\documentclass[12pt]{article}
\usepackage{pic02}
\usepackage{hyperref}
\usepackage{url}
\usepackage{graphicx}
\usepackage{epsfig}

\def\xp{$x^{\prime}$} 
\def\yp{$y^{\prime}$}

\def\dz{$D^{0}$}
\def\dzb{$\overline{D^{0}}$}
\def\dplus{$D^{+}$}

\def\dzdzb{$D^{0}-\overline{D^{0}}$}
\def\bzbzb{$B^{0}-\overline{B^{0}}$}
\def\kzkzb{$K^{0}-\overline{K^{0}}$}
\def\dztokspp{$D^{0}\rightarrow K^{0}_{S}\pi^{+}\pi^{-}$}

\def\dztokpws{$D^{0}\rightarrow K^{+}\pi^{-}$}
\def\dztokprs{$D^{0}\rightarrow K^{-}\pi^{+}$}
\def\dztokppzws{$D^{0}\rightarrow K^{+}\pi^{-}\pi^{0}$}

\def\dztokpppws{$D^{0}\rightarrow K^{+}\pi^{-}\pi^{+}\pi^{-}$}

\def\dztokzpz{$D^{0}\rightarrow K^{0}\pi^{0}$} 

\def\dztokzspz{$D^{0}\rightarrow K^{0}_{S}\pi^{0}$} 
\def\dztokzlpz{$D^{0}\rightarrow K^{0}_{L}\pi^{0}$} 
\def\dztokk{$D^{0}\rightarrow K^{+}K^{-}$}
\def\dztopp{$D^{0}\rightarrow \pi^{+}\pi^{-}$}
\def\dztokmunu{$D^{0}\rightarrow K^{+}\mu^{-}\overline{\nu_{\mu}}$} 
\def\dztokstenu{$D^{0}\rightarrow K^{*+}e^{-}\overline{\nu_{e}}$}

\begin{document}

\title{\bf RECENT ADVANCES IN CHARM PHYSICS}
\author{
Alex Smith       \\
{\em University of Minnesota} \\
{\em 116 Church St. S.E.} \\
{\em Minneapolis, MN 55455}}
\maketitle

%
%
\begin{figure}[h]
\begin{center}
%
%
%
%
%
\end{center}
\end{figure}

\baselineskip=14.5pt
\begin{abstract}
New results from charm experiments have led to renewed interest
in this physics.  The charm sector is now seen as 
a powerful tool to search for new physics and to advance our understanding
of the standard model.  We owe much of this progress to the combination
of precision vertexing and large data samples 
collected by recent $e^{+}e^{-}$ and fixed target experiments.
Sensitivities to \dzdzb\ 
mixing and $CP$ violation are approaching some
non-standard model predictions.  Recent measurements of charmed
particle lifetimes and semileptonic decays  have added to our
understanding of decay mechanisms and the dynamics of heavy-to-light
quark transitions.  Many of these
provide vital input to QCD models and are an essential ingredient in
extracting standard model parameters from other measurements.  
Studies of charmonium production continue to offer new surprises. 
A recent measurement from Belle indicates that $\sim 60\%$ of $J/\psi$
events produced in continuum $e^{+}e^{-}$ collisions are produced with an
additional charm quark pair.  Fueled by new data from a host of continuing
and future experiments, we can expect significant improvement to the
standard model and possibly some new surprises.
\end{abstract}
\newpage

\baselineskip=17pt


\section{Why Charm Physics?}

Until recently charm physics has often been overlooked as a tool to
understand the standard model and to search for physics beyond the
standard model.  The 
reason for this is that the charm quark is neither
heavy nor light enough to apply the approximations that have been
successful in modeling the dynamics of bottom and strange particles. 
Long range effects may spoil perturbative predictions of charm decay
rates.  Advances in lattice QCD, recent advances in experimental
sensitivities, and the promise of dramatic improvements from current and
future charm experiments has revived interest in this field.

Measurements of several important standard model parameters are 
presently limited by theoretical uncertainties of
non-perturbative QCD.  All decay processes involving
hadrons are 
modified by soft processes.  The charm sector provides stringent tests of
non-perturbative QCD.  Recent advances in lattice QCD calculations have
allowed 
predictions to a few percent accuracy for several ``gold-plated''
calculations, including 
many charmed particle masses, decay constants, semileptonic form
factors~\cite{Lepage}. 
Accurate measurements of these quantities will provide crucial tests of the
${\cal O}(1\%)$ uncertainties claimed by lattice QCD.  

It is also important to understand the relative importance of different
processes which play a role in production and decay of charmed particles.
The mechanisms for production of charmonium in
$p\overline{p}$ collisions are still not understood.  Recent
measurements from the $B$ factories have further challenged our
understanding of these mechanisms.  Measurements of charmed meson and
baryon lifetimes provide vital information about the relative importance
of different decay processes.  

There is potential to observe new physics in the charm sector through
searches for \dzdzb\ mixing and $CP$ violation.  Until recent advances
in sensitivity, limits on \dzdzb\ mixing and $CP$ violation in many
modes were too large to be of interest.  New measurements have enabled
the search for new physics to extend into previously inaccessible
corners.  For example, \dzdzb\ mixing would be sensitive to 
down-type non-standard model particles, which may appear in the box
diagram for mixing.  Such particles would not be observable through 
\bzbzb\ and \kzkzb\ mixing, which are only sensitive to up-type particles. 

The standard model contributions to $CP$ violation and \dzdzb\ mixing in
charm decays are expected to be quite small.  There is uncertainty in
the magnitude of enhancements due to long-distance effects, but these
are believed to be well below the present experimental sensitivities.
Furthermore, experimental measurements may shed light on the magnitude
of these long-distance effects.  

There are several important topics not covered in this summary due to time and
space constraints such as rare and forbidden decays, charmonium and
charmed baryon spectroscopy, measurement of the $D^{*+}$ width, and
tests of $CPT$ invariance, to
name a few.  The use of Dalitz analyses of multi-body  
charm meson decays and radiative $J/\psi$ decays to study light
meson/glueball/exotic spectroscopy is covered in other talks from this
conference by Brian Meadows and Shen Xiaoyan. 


\section{Searches for New Physics Using D Meson Decays}

\subsection{\dzdzb\ Mixing} 

\dzdzb\ mixing is described by amplitudes $x \equiv \Delta M/\Gamma_{D}$ and $y
\equiv\Delta \Gamma/2\Gamma_{D}$, where $x$ and $y$ arise from
differences in the masses ($\Delta M$) and widths ($\Delta \Gamma$),
respectively, of the mass eigenstates of the \dz\ meson, where $\Gamma_{D}$
is the observed 
width of the $D^{0}$ meson.  The standard model predictions for $x$ and
$y$ are below $10^{-3}$.  The amplitude $x$ could be further suppressed by the
GIM\cite{GIM} mechanism, however both $x$ and $y$ could be enhanced by
long-distance contributions.  
New physics could lead to an enhancement of $x$, but is not expected to
contribute to the amplitude $y$.

One searches for \dzdzb\ mixing by studying the ``wrong-signed'' (WS)
final state of \dz\ meson decays, such as \dztokpws~\cite{chargeconj}.
Contributions to the WS signal may come from the \dz\ mixing into a
\dzb\ followed by a Cabibbo-favored (CF) decay or from standard
model doubly Cabibbo-suppressed (DCS) decays with amplitude $R_{D}$ in
the case of hadronic final states.  The ``right-signed'' (RS)
decays come from Cabibbo-favored decays, such as \dztokprs.   

In order to extract mixing parameters from hadronic final states, such
as \dztokpws, one must use reconstructed \dz\ candidate proper time
information to distinguish possible contributions from $x$, $y$, and
$R_{D}$.  The time dependence of the
amplitude contains a pure DCS term, a pure mixing term, and an
interference term, each with a distinct proper time distribution:
\begin{equation}  
\label{eq:mixtimedep}
r(t)=\left[ R_{D} + \sqrt{R_{D}}y^{\prime}t 
+ \frac{1}{4}(x^{\prime 2} + y^{\prime 2})t^{2} \right] e^{-t}.
\end{equation}
The primes on the $x$ and $y$ indicate that there may be a strong phase
difference, $\delta_{\rm fs}$,  between the DCS and CF decays in decays
to hadronic final states, which modifies
the values $x$ and $y$ and depends on the final state under
consideration.  The two are related by a rotation:
$x^{\prime}=x\cos\delta_{\rm fs} + y\sin\delta_{\rm fs}$ and
$y^{\prime}=y\cos\delta_{\rm fs} - x\sin\delta_{\rm fs}$.
Since this strong phase difference is difficult to estimate from
calculations, it is {\em essential} that it measured 
experimentally in order to distinguish $x$ and $y$~\cite{Petrov}.

For semileptonic \dz\ decays there are no DCS contributions ($R_{D}=0$)
and Eq.~\ref{eq:mixtimedep} simplifies to  
\begin{equation}  
\label{eq:slmixtimedep}
r(t)=\frac{1}{4}(x^{\prime 2} + y^{\prime 2})t^{2}e^{-t}.
\end{equation}
Observation of a WS signal in a semileptonic mode would be an indication
of mixing, however, due to the absence of the interference term, one
cannot distinguish whether this mixing is from $x$ or $y$.

\begin{figure}[htb]
  \begin{center}
    \includegraphics[width=12.5cm]{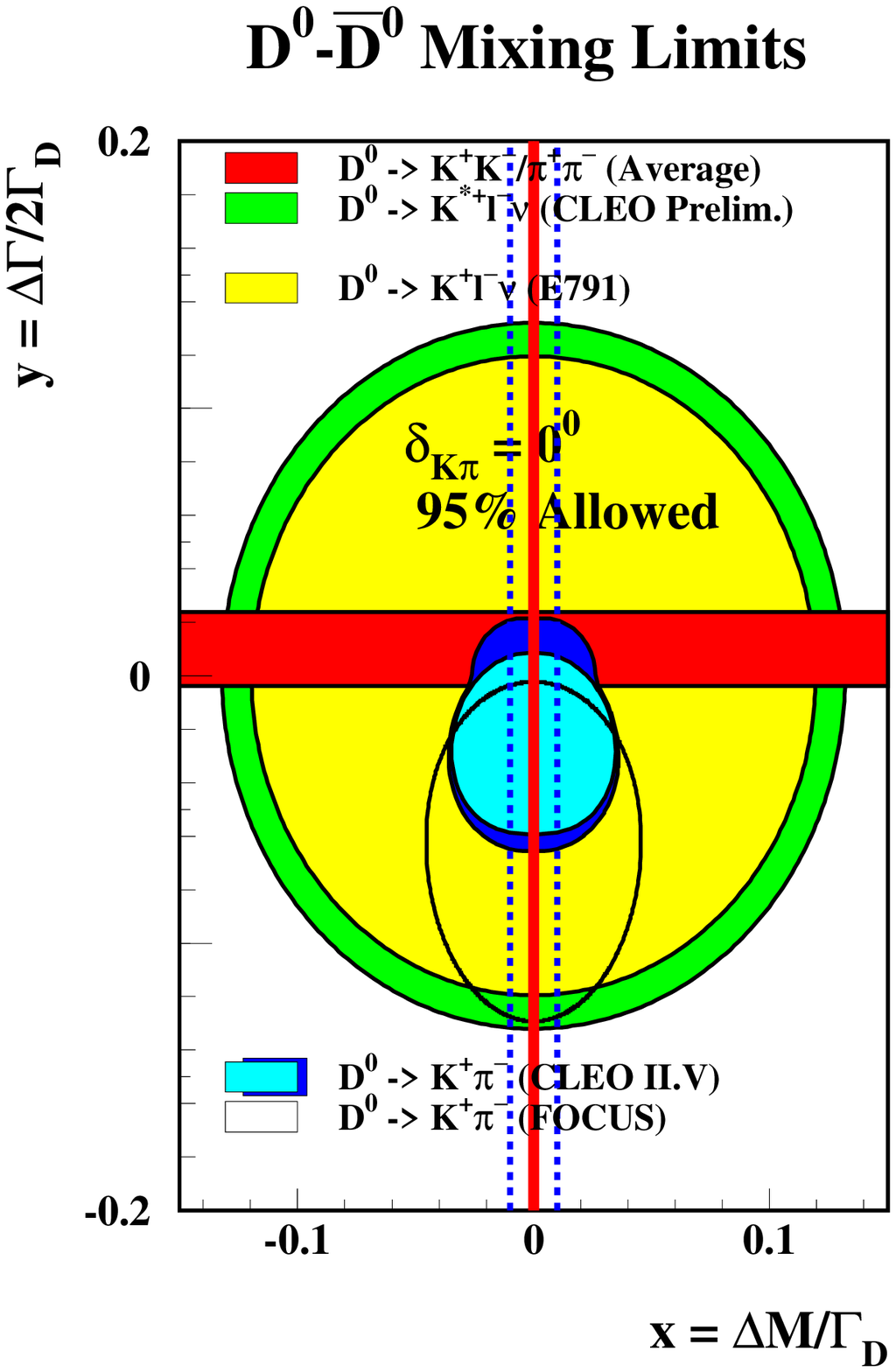}
    \end{center}
 \caption{
      Our present knowledge of \dzdzb\ mixing.  The solid vertical lines
      indicate a ``typical'' standard model prediction for $x$.  The
      dashed vertical lines indicate the upper range of non-standard
      model predictions for $x$.  The strong phase shift
      $\delta_{\rm fs}$ between the Cabibbo-favored and DCS decays is assumed
      to be zero in plotting the \dztokpws\ results.  While the strong
      phase shift is expected to be close to zero, until it is actually
      measured, the allowed region from the \dztokpws\  
      measurements must be expanded to include the area swept out by
      rotating these regions about the origin.
    \label{dmixfig} }
\end{figure}

The present limits on \dzdzb\ mixing parameters $x$ and $y$ from
different measurements are summarized in Fig.~\ref{dmixfig}.  The limits
from the channel \dztokpws\ are plotted assuming $\delta_{\rm fs}=0$.  If
this assumption is not made, all possible rotations of these regions
about the origin must be considered, leading to much weaker limits.

\subsubsection{\dztokpws} 

Presently, the best limits on \xp\ come from measurements of
\dztokpws\ from the CLEO~\cite{CLEOkp} and 
FOCUS~\cite{FOCUSkp} collaborations.  CLEO performs fits
with and without the assumption of $CP$ conservation.  Both
experiments observe WS signals consistent with DCS, but both seem to
favor negative values of \yp.  The 95\%
confidence limits on mixing parameters \xp\ and  \yp\
from CLEO,  assuming $CP$ conservation, are $|x^{\prime}|<2.8\%$ and
$-5.2\%<y^{\prime}<0.2\%$, respectively.  The corresponding limits from
FOCUS are $|x^{\prime}|<3.9\%$ and $-12.4\%<y^{\prime}<-0.6\%$.   

Both Belle and BaBar have preliminary measurements of the time-integrated WS
rate, $R_{WS}$ based on $317$ and $210$ events respectively,
corresponding to WS rates of $R_{WS}=(0.372 \pm 0.025
^{+0.009}_{-0.014})\%$\cite{bellekp} and   
$R_{WS}=(0.383 \pm 0.044 \pm 0.022)\%$~\cite{babarkp, uncert}.  
The combined world average is $R_{WS}=(0.37
\pm 0.02)\%$~\cite{myaverages}.    

\subsubsection{$D^{0}$ Decays to Multi-body Final States} 

CLEO has also measured $R_{WS}$ in multi-body channels \dztokppzws\ and
\dztokpppws\ with results
$R_{WS}=(0.43^{+0.11}_{-0.10}\pm 0.07)\%$ and
$R_{WS}=(0.41^{+0.12}_{-0.11} \pm 0.04)\%$,
respectively~\cite{CLEOKppz,CLEOKppp}.  $R_{WS}$ need not be the same
for different decay modes, however.  With the large data samples from
the $B$ factories, it may be possible to set \dzdzb\ mixing limits using
combined Dalitz plot and proper time fits in multi-body modes.  These
modes may prove useful in searching for $CP$ violation and understanding
strong phase shifts~\cite{Rosner}.

\subsubsection{$\delta_{K\pi}$ from \dztokzpz} 

Measurements of \dztokpws\ cannot distinguish between $x$ and $y$
without knowledge of the strong phase shift $\delta_{K\pi}$ between the
CF and DCS decays. 
Furthermore, if $y$ turns out to be large, then it will not be possible
to extract precise limits on $x$ without knowledge of $\delta_{K\pi}$.
While the theoretical bias is toward a small phase, this quantity must
be measured in order to distinguish between new physics and $y$-type standard
model mixing.  The strong phase $\delta_{K\pi}$ can be pinned down by 
comparing different $D\rightarrow K\pi$ decay rates. 
Belle has measured the asymmetry of the rates of \dztokzspz\ and the previously
unmeasured \dztokzlpz~\cite{bellekzpz, belledpf}.  Their preliminary 
measurement is not yet sensitive enough to provide a measurement of
$\delta_{K\pi}$ 
\begin{equation}
A=\frac{\Gamma(D^{0}\rightarrow K^{0}_{S}\pi^{0}) -
  \Gamma(D^{0}\rightarrow K^{0}_{L}\pi^{0})} 
{\Gamma(D^{0}\rightarrow K^{0}_{S}\pi^{0}) + \Gamma(D^{0}\rightarrow
  K^{0}_{L}\pi^{0})} = 0.06 \pm 0.05 \pm 0.05 ,
\end{equation}  
however, improvements are expected as more data comes in.

\subsubsection{\dztokspp} 

The decay \dztokspp\ may be used to measure $x$ and $y$ directly since
the strong phase difference may be extracted simultaneously in a
time-dependent fit to the Dalitz plot.  This is possible because both
the RS and WS decays in the submode $D^{0}\rightarrow
K^{*\pm}\pi^{\mp}$ have the same final state.  Thus, one can fit for the phase
difference directly.  The sign of $x$ can also be extracted from such a
fit.

CLEO has presented evidence for a WS amplitude and has measured the
branching fraction relative to the RS mode to be 
\begin{equation}
\frac{ {\cal B}(D^{0}\rightarrow K^{*+}\pi^{-}) }
{ {\cal B}(D^{0}\rightarrow K^{*-}\pi^{+}) } 
= ( 0.5\pm 0.2\ ^{+0.5}_{-0.1}\ ^{+0.4}_{-0.1} )\% 
\end{equation}
and the 
strong phase difference between the RS and WS to be $(189^{\circ} \pm
10 \pm 3 ^{+15}_{-5})^{\circ}$ using a time-independent Dalitz plot
fit~\cite{asnerkzpp}.  The last uncertainty is due to the choice of
resonances and model.  No $CP$ violating effects were observed when
separating the sample into \dz\ and \dzb\ subsamples.
Results of a 
time-dependent fit with limits on $x$, $y$, and $CP$ violation are
expected soon.  This channel may offer the greatest sensitivity to
$x$ at the large integrated luminosities already collected by Belle and
BaBar.

\subsubsection{Measurement of $y_{CP}$ Using \dztokk\ and \dztopp } 

The decays of the \dz\ to $CP$ eigenstates may be used to measure the
amplitude 
\begin{equation}
y=\frac{\Delta\Gamma}{2\Gamma}=\frac{\Gamma_{CP^{+}}-\Gamma_{CP^{-}}}
{\Gamma_{CP^{+}}+\Gamma_{CP^{-}}}.
\end{equation}
Decays to the $CP$-even final states $K^{+}K^{-}$ and $\pi^{+}\pi^{-}$
are the most accessible experimentally and have been studied by
BaBar~\cite{babarkk}, Belle~\cite{bellekk}, CLEO~\cite{cleokk},
E791~\cite{E791kk}, and FOCUS~\cite{focuskk}.  The value of $y$ is
extracted by assuming 
$CP$ conservation and comparing with the well-measured lifetime of the
non-$CP$ \dztokprs\ decay mode:
\begin{equation}
y_{CP}=\frac{\tau_{K^{-}\pi^{+}}}
{\tau_{K^{+}K^{-}}} - 1 .
\end{equation}
Many systematic uncertainties cancel in this ratio.

The present world average of $(1.0\pm 0.7)\%$~\cite{myaverages} is
dominated by recent 
measurements from the BaBar, Belle, and FOCUS experiments of  
$(1.4 \pm 1.0^{+0.6}_{-0.7})\%$~\cite{babarkk}, $(-0.5 \pm
1.0\pm0.8)\%$~\cite{bellekk}, and $(3.4 \pm 1.4 \pm
0.7)\%$~\cite{focuskk}, respectively.

\subsubsection{Semileptonic Channels} 

As shown in Eq.~(\ref{eq:slmixtimedep}), a signal for WS semileptonic 
\dz\ decay would be evidence for $x$- or $y$-type \dzdzb\ mixing.  No
signal has been observed and upper limits have been set.
The most recent limits on $R_{mix}\equiv 1/2(x^{2}+y^{2})$ using
semileptonic WS decays are 
from E791 ($<0.5\%$ @ 90\%C.L.)~\cite{E791sl} and CLEO ($<0.87\%$ @
95\%C.L.)~\cite{CLEOsl}, measured in the  
channels \dztokmunu\ and \dztokstenu, respectively.  When comparing,
note that the E791 measurement is quoted as a 90\% C.L. limit.   
FOCUS has presented an {\em estimated} sensitivity of $<0.12\%$ @
95\%C.L. (with statistical errors only) in the channel
\dztokmunu~\cite{FOCUSichep} but has not yet presented an actual result.

\subsection{$CP$ Violation}

$CP$ violation may manifest itself in three possible ways:  1) as a
difference in the decay rates for charge conjugate states ($D \rightarrow
f \neq \overline{D} \rightarrow \overline{f}$), 2) as an asymmetry in
the mixing rate for charge conjugate states ($D^{0}\rightarrow
\overline{D^{0}} \neq \overline{D^{0}} \rightarrow 
D^{0}$), or 3) as a difference in the phase of interference between
mixing and decay contributions for charge conjugate states.

Two ingredients are required in order to observe non-standard model
physics through $CP$ violation.  First, the decay amplitude must have
contributions from at least two diagrams with different weak phases.
Second, there must be a non-negligible strong phase shift between two of
the processes.  The strong phase shift is expected to be non-zero in charm
decays, since $SU(3)$ flavor symmetry is known to be badly violated in
some decays.  For example, the ratio of rates 
$BR(D^{0}\rightarrow K^{+}K^{-})/BR(D^{0}\rightarrow \pi^{+}\pi^{-})$
is approximately three times larger than the value of one predicted
under the assumption of $SU(3)$ flavor symmetry.

Since $CP$ violation is expected to be zero for doubly Cabibbo-suppressed and
Cabibbo-favored decays within the standard model this is a good place to
look for new physics.  Many doubly Cabibbo-suppressed modes were
observed only recently and are being
studied for the first time. 
CLEO has measured $CP$ asymmetries in decay, mixing, and interference in 
the doubly Cabibbo-suppressed channel \dztokpws\ to be consistent with
zero: $A_{\rm decay}=(-1^{+16}_{-17}\pm 1)\%$, 
$A_{\rm mixing}=(23^{+63}_{-80}\pm 1)\%$, $\sin\theta=(0 \pm 60 \pm
1)\%$~\cite{CLEOkp}.  

For singly Cabibbo-suppressed decays, the standard model prediction is
of order $10^{-3}$ or below, and arises out of the interference between
tree and penguin amplitudes.   Asymmetries in several modes have been
measured recently by CLEO~\cite{cleokkpp}, FOCUS~\cite{focuskk}, and
E791~\cite{E791kk}, including \dztokk, \dztopp, $D^{0}\rightarrow
K^{+}K^{-}\pi^{+}$ channels, which are now measured to be $(0.48\pm
1.57)\%$,  
$(2.1\pm 2.6)\%$, and $(0.2\pm 1.1)\%$, respectively~\cite{myaverages}.

Several other decay channels have also been studied and are not covered
in this paper.  We can look forward to $CP$ violation searches utilizing
not only comparisons of rates, but also the detailed amplitude and phase
observables from Dalitz plot analyses of multi-body modes.


\section{Measurements Which Provide Input to QCD Models}

\subsection{Charm Semileptonic Decays}

Our understanding of many important standard model parameters, 
such as the Cabibbo-Kobayashi-Maskawa (CKM) matrix elements $|V_{ub}|$
and $|V_{cb}|$ are limited by theoretical uncertainties in the QCD 
of heavy quark semileptonic decays.  Many of these
uncertainties in $B$ meson decays can be reduced using lattice QCD with 
input from analogous charm decays, such as 
$D^{+}\rightarrow \overline{K^{*0}}\ell^{+}\nu_{\ell}$,   
$D^{0}\rightarrow \pi\ell\nu_{\ell}$, $\rho\ell\nu_{\ell}$, 
$K\ell\nu_{\ell}$ or $D_{s}^{+}\rightarrow \phi\ell\nu_{\ell}$.

The normalized branching fraction of the decay 
$D^{+}\rightarrow \overline{K^{*0}}\ell^{+}\nu_{\ell}$ 
\begin{equation}
  R_{\ell}^{+} \equiv \frac
  {\Gamma(D^{+}\rightarrow \overline{K^{*0}}\ell^{+}\nu_{\ell})}
  {\Gamma(D^{+}\rightarrow \overline{K^{-}}\pi^{+}\pi^{+})}
\end{equation}
 has been 
measured most recently by the 
FOCUS $(R_{\mu}^{+}=0.602 \pm 0.010 \pm 0.021)$~\cite{FOCUSDplussl}, 
CLEO $(R_{e}^{+}: 0.74 \pm 0.04 \pm 0.05,\ 
R_{\mu}^{+}: 0.756 \pm 0.105 \pm 0.06)$)~\cite{cleoDplussl}, 
E687 $(R_{\mu}^{+}=0.588 \pm 0.042 \pm 0.063)$~\cite{PDG}, and  
E691 $(R_{e}^{+}=0.49 \pm 0.04 \pm 0.05)$~\cite{PDG}
experiments. 
FOCUS observed dramatic interference effects in this decay, which result
in a 
large asymmetry in the $\overline{K^{*}}$ decay angle for masses below
the $\overline{K^{*}}$ pole mass and almost no asymmetry for masses 
above the pole~\cite{FOCUSDplussl}.  They find this to be consistent with a small, but 
significant even-spin contribution to the $K\pi$ final state.  They also
measure the 
semileptonic form factor ratios $r_{v}$ and $r_{2}$ including the
$S$-wave component to be $1.504 \pm 0.057 \pm 0.039$ and $0.875 \pm
0.049 \pm 0.064$, respectively.
 
FOCUS also has a preliminary measurement of $D_{s}^{+}\rightarrow
\phi\mu^{+}\nu_{mu}$ with a branching fraction of 
$0.54 \pm 0.033 \pm 0.048$~\cite{FOCUSichep}.

\subsection{Charmed Meson and Baryon Lifetimes}

Measurements of charmed baryon and meson lifetimes provide important
insight into the decay processes of heavy mesons and baryons.
Depending on the particular decay, different processes such as
external spectator, internal spectator, $W$ exchange, or annihilation
may be important.  Interference effects may have a large role in 
determining the observed lifetime.  Comparisons of non-perturbative QCD
models with the measured lifetime hierarchy of charmed particles provide
an important test of these models.  

The striking difference in the observed \dz\ and \dplus\ lifetimes is
now understood to be due to interference between diagrams contributing to
\dplus\ decays, but not those contributing to \dz\ decays. 
The $D^{0}$ and $D^{+}$ lifetimes are now measured to a fraction of a
percent, $410.4 \pm 1.5$~fs~\cite{myaverages, FOCUSDlife, E791kk,
SELEXLambdaDzero,
  CLEODlife, PDG} and  $1042.7 \pm
6.9$~fs~\cite{myaverages, FOCUSDlife, CLEODlife, PDG}, 
respectively, and the $D_{s}^{+}$ lifetime is measured to about two
percent, $490.7\pm 8.4$~fs~\cite{myaverages, SELEXDs, PDG}.  The
$\tau_{D^{0}}$ and $\tau_{D^{+}}$ averages are
dominated by recent measurements from 
the FOCUS collaboration~\cite{FOCUSDlife, PDG} of $409.6\pm 1.1 \pm 1.5$~fs
and  $1039.4\pm 4.3 \pm 7.0$~fs, respectively.  A preliminary
measurement of the $D^{+}_{s}$ lifetime from FOCUS of $506\pm 8$~fs 
using half of their data sample was not included in the average, since
the systematic uncertainty was not yet known.

Lifetimes of the $\Lambda_{c}^{+}$, $\Xi_{c}^{+}$, and $\Xi_{c}^{0}$,
charmed baryons have been measured recently by the fixed target
experiments FOCUS~\cite{FOCUSbaryon},
SELEX~\cite{SELEXLambdaDzero}, and 
E687\cite{PDG} and by the  
CLEO $e^{+}e^{-}$ experiment~\cite{CLEObaryon}.  
Most of the charmed baryon lifetime hierarchy is  
described quite well by the theory.  One notable exception is the
ratio of the lifetimes of $\Xi_{c}^{+}$ and $\Lambda_{c}^{+}$, for which
the measurement is approximately a factor of two larger than the
prediction.  New and more precise measurements of charmed hadron 
lifetimes and decay modes will provide important guidance to 
our understanding of heavy hadron decays.

\subsection{Charmonium Production Mechanisms}

Measurements of charmonium production have provided many surprises
and many challenges to our conception of how such particles are produced.
During Run~I of the Tevatron, the production cross sections for
charmonium and bottomonium states exceeded NRQCD predictions by as much
as two orders of magnitude~\cite{D0psi, CDFpsi}.  Many new contributing
processes were proposed, including new fragmentation contributions 
and production in a color-octet state~\cite{BraatenCOct}.  

Recently, the Belle and BaBar collaborations have made measurements
which test NRQCD using $e^{+}e^{-}$ collisions below the $\Upsilon(4S)$.
At these energies, the following processes are expected to contribute:
\begin{flushleft}
\begin{eqnarray}
e^{+}e^{-} \rightarrow J/\psi gg & {\rm Singlet, octet} & {\rm dominant} \label{eq:dom} \\
e^{+}e^{-} \rightarrow J/\psi g  & {\rm Octet} & {\rm dominant\ at\
  endpoint}  \label{eq:CO} \\ 
e^{+}e^{-} \rightarrow J/\psi c\overline{c} & {\rm Octet,\ singlet}
&{\rm Four\ charm\ gluon\ 
  splitting--\ {\cal O}(10\%) }  \label{eq:fourc} \\
e^{+}e^{-} \rightarrow J/\psi q\overline{q} & {\rm Octet} & {\rm Two\ charm\ gluon\
splitting--\  Small} 
  \label{eq:small}
\end{eqnarray}
\end{flushleft}

Both the BaBar and Belle collaborations have observed $J/\psi$
production in 
the continuum below the $\Upsilon(4S)$ resonance.  One may test for the
color-octet contribution of Eq.~(\ref{eq:CO}) predicted by NRQCD by
examining the momentum $p^{*}$ and polar angle ($\theta^{*}$) of the
$J/\psi$ in the center-of-mass frame.  The angular
$\cos\theta^{*}$ distribution may be fit to $1+A\cdot\cos^{2}\theta^{*}$
in low and high $p^{*}$ bins in order to test the models.
NRQCD and color singlet models both predict a flat ($A=0$) distribution
at low $p^{*}$.  At high momentum the color singlet model predicts 
$A \sim -0.8$ while NRQCD predicts $0.6 < A < 1.0$.  BaBar has performed
these fits for $p^{*}<3.5$~GeV/$c$ and $p^{*}>3.5$~GeV/$c$ and find
$A=0.05 \pm 0.22$ and $A=1.5\pm 0.6$,
respectively~\cite{babarpromcharmonium}, which is consistent with NRQCD.
The same measurement from Belle~\cite{bellepromcharmonium} yields a
large positive value $A=0.9\pm 0.2$ at all momenta.  This distribution
is only expected for the color singlet four-charm gluon splitting of
Eq.~(\ref{eq:fourc}), which is predicted to be small.  In some models,
the leading 
color-octet mechanism of Eq.~(\ref{eq:CO}) is expected to contribute
only in the high $p^{*}$ end-point region, where it would give
$A\sim +1$.  No excess was observed in the high $p^{*}$ region 
in either measurement.

Belle recently presented a surprising result indicating
that four-charm production (Eq.~(\ref{eq:fourc})) comprises the 
{\em majority} of continuum $J/\psi$ production~\cite{belledpf}.
They studied the spectra of mass recoiling against the $J/\psi$ and
observe a clear threshold at twice the charm mass and evidence for 
peaks at the $\eta_{c}$, $\chi_{c0}$, and $\eta_{c}^{\prime}$ masses.
They also search for a third associated charm quark through the decays
$e^{+}e^{-}\rightarrow J/\psi D^{*}X$ and 
$e^{+}e^{-}\rightarrow J/\psi D^{0}X$.  They observe signals of 5.3 and 
3.7 standard deviations statistical significance.
Using the JETSET
fragmentation rates they convert these rates into a cross section for 
$e^{+}e^{-}\rightarrow J/\psi c\overline{c}$.  They find that four-charm
production accounts for approximately $60\%$ of continuum $J/\psi$
production:
\begin{equation}
\frac{\sigma(e^{+}e^{-}\rightarrow J/\psi c\overline{c})}
{\sigma(e^{+}e^{-}\rightarrow J/\psi X)} = 0.59 ^{+0.15}_{-0.13} \pm
0.12 .
\end{equation}
This is quite a surprise considering that the prediction is of order
$10\%$.  As more data comes in from both experiments, our understanding
of these production mechanisms should become more clear.


\section{The Future of Charm Physics}

FOCUS, CLEO, and E791 continue to produce important results using their
well-developed analysis tools and final data samples.
Many of the most challenging analyses involving Dalitz plot and proper
time fits are now bearing physics results.   

The B factory experiments 
Belle and BaBar are expected to add approximately a factor of ten to
their already large data samples over 
the next few years.  Many of the
first round of analyses from these experiments are close to bearing
results with a factor of 5-10 times the CLEO statistics.  

Dedicated charm experiments CLEO-c/CESR-c and BES~III are proposed for 2003 and
2005/6, respectively.  A funding decision will be made soon regarding
the CLEO-c/CESR-c proposal.  CLEO-c will allow precision
measurement of the decay constants $f_{D}$ and $f_{D_{s}}$ to a
precision of $2.3\%$ and $1.7\%$, respectively, using a sample of
approximately 30 million events (six million {\em tagged} $D$ decays)--
310 times the Mark III data sample.  Precise measurements of several
important absolute branching fractions, semileptonic form factors will
be made, as well as high statistics searches for \dzdzb\ mixing, $CP$
violation, and rare $D$ decays.

CDF has demonstrated the ability to trigger on charm decays in the messy
environment of $p\overline{p}$ collisions using its
silicon trigger.  D\O\ is planning to implement a silicon vertex trigger
during Run~II of the Tevatron.  These experiments benefit from a large
charm cross section which is approximately a factor of ten larger than
the bottom cross section.  Extrapolating from the preliminary CDF
results~\cite{CDFcharm} and assuming that the trigger rates are
sustainable at higher
luminosities, one can expect approximately $10^{7}$ \dztokprs\ events.
Assuming the same efficiency ratio for WS, one expects approximately
15,000 WS \dztokpws\ events and a $CP$ violation reach of perhaps
$10^{-3}$.  

Toward the end of this decade, the proposed BTeV and LHCb experiments
are expected to take data.  The BTeV trigger will require only two tracks
with a detached vertex and 
will have a large acceptance for charm.  The LHCb trigger is not
expected to have significant acceptance for charm.  Using very crude
estimates and 
many assumptions, one expects approximately $10^{8}$ RS \dztokprs\
events and approximately 150,000 WS \dztokpws\ events to be collected by
BTeV.  Such samples
would allow a $CP$ violation reach down to $\sim 10^{-4}$.


\section{Summary}

Charm physics has proved to be an important tool for understanding the
standard model and searching for new physics.  Recent searches for new
physics are starting to exclude some non-standard model predictions.  
Lattice QCD awaits validation of its predictions of form factors,
rates, and decay constants of charm decays.  These measurements effect
the determination of other important quantities, such as $|V_{ub}|$ and
$|V_{cb}|$.  Similarly, studies of charmed hadron lifetimes feed back
into our understanding of decay processes.  Finally, measurements of 
charmonium production, both at hadron and $e^{+}e^{-}$ colliders
continue to offer new surprises and challenges to NRQCD.  The interest
in charm physics will continue to grow as the $\Upsilon(4S)$, charm 
factories, and hadron experiments weigh in with new and more precise
measurements.


\begin{thebibliography}{99}
\bibitem{Lepage} G. P. Lepage, High Precision Nonperturbative QCD,
 Presentation at Hadron Physics 2002, April 2002, Bento Goncalves, Brazil. 
\bibitem{GIM} S. L. Glashow, J. Iliopolous, and L. Maiani, Phys. Red. D
  {\bf2}, 1285 (1970).
\bibitem{chargeconj} Charge conjugate modes are implied throughout this
  paper except where explicitly noted.
\bibitem{Petrov} A. F. Falk, Y. Nir, A. A. Petrov, JHEP 9912 (1999) 019.
\bibitem{CLEOkp} R. Godang {\em et al.}, Phys. Rev. Lett. {\bf 84}, 5038
  (2000). 
\bibitem{FOCUSkp} J. M. Link {\em et al.}, Phys. Rev. Lett. {\bf 86},
  2955 (2001); J. M. Link, Recent Results in Charm Mesons,
  Presentation at The Fifth KEK Topical Conference-- Frontiers in Flavor
  Physics, November, 2001, Tsukuba, Japan.
\bibitem{bellekp} K. Abe, {\em et al.}, ``A Measurement of the Rate of
  Wrong-sign Decays $D^{0}\rightarrow K^{+}\pi^{-}$'', Proceedings of
  The 31st International Conference on High Energy Physics, Amsterdam,
  July 2002, hep-ex/0208051.
\bibitem{babarkp} U. Egede, ``Determination of the Wrong Sign Decay Rate
  $D^{0}\rightarrow K^{+}\pi^{-}$ and the Sensitivity to \dzdzb\
  Mixing'', Proceedings of the International Europhysics
  Conference on HEP, Budapest, July, 2001, hep-ex/0111062. 
\bibitem{uncert} Unless otherwise noted, the statistical uncertainty is
  quoted first and the systematic is quoted second.  For averages, the
  systematic and statistical uncertainties are added in quadrature.   
\bibitem{myaverages} These are my weighted averages including the 
  Particle Data Group measurements and all known published or preliminary 
  measurements up to and including those presented at the ICHEP 2002 
  conference.  Averages were calculated using the technique summarized
  in the Review of Particle Physics. 
\bibitem{CLEOKppz} G. Brandenburg {\em et al.}, Phys. Rev. Lett. {\bf
  87}, 071802 (2001). 
\bibitem{CLEOKppp} S. A. Dytman {\em et al.}, Phys. Rev. D {\bf 64},
  111101 (2001). 
\bibitem{Rosner}  C. Chiang, J. L. Rosner,  Phys. Rev. D {\bf 65}, 
 054007 (2002).
\bibitem{bellekzpz} K. Abe {\em et al.}, ``Measurement of $D^0$ decays
  to $K^0_L \pi^0$ and $K^0_S \pi^0$ at Belle'', XX International
  Symposium on Lepton and Photon Interactions at High Energies,
  Rome, July 2001, hep-ex/0107078.
\bibitem{belledpf}   B. Yabsley, ``Current Charm Studies at Belle'',
  presented at the Meeting of the Division of Particles and Fields,
  Williamsburg, May 2002. 
\bibitem{asnerkzpp} H. Muramatsu, {\em et al.}, submitted to
  Phys. Rev. Lett., hep-ex/0207067. 
\bibitem{babarkk} A. Pompili, ``Charm Mixing and Lifetimes at BaBar'', 
XXXVII Rencontres de Moriond on Electroweak Interactions and Unified
Theories, Les Arcs, March 2002, hep-ex/0205071. 
\bibitem{bellekk} K. Abe {\em et al.}, Phys. Rev. Lett. {\bf 88}, 162001
  (2002). 
\bibitem{cleokk} S. E. Csorna {\em et al.}, Phys. Rev. D {\bf 65}, 092001
  (2002).
\bibitem{E791kk} E. M. Aitala {\em et al.}, Phys. Rev. Lett. {\bf 83}, 32 (1999).
\bibitem{focuskk} J. M. Link {\em et al}, Phys. Lett. B {\bf 485}, 62
  (2000). 
\bibitem{E791sl} E. M. Aitala {\em et al}, Phys. Rev. Lett. {\bf 77}, 2384 (1996)
\bibitem{CLEOsl} A. Smith, ``Recent Charm Results from CLEO'',
  presented at XXXVIIth Recontres de Moriond QCD, Les Arces, March 2002.
\bibitem{FOCUSichep} S. Malvezzi, ``Charmed Mesons Lifetimes, Decays,
  Mixing and CPV Results from FOCUS'', 31st International Conference on  
  High Energy Physics, Amsterdam, July 2002. 
\bibitem{cleokkpp} S. E. Csorna {\em et al.}, Phys. Rev. D {\bf 65},
  092001 (2002). 
\bibitem{FOCUSDplussl} J. M. Link {\em et al}, Phys. Lett. B {\bf 541},
  243 (2002); J. M. Link {\em et al.}, Phys. Lett. B {\bf535},
  43 (2002); J. M. Link {\em et al.}, hep-ex/0207049 (2002);
\bibitem{cleoDplussl} G. Brandenburg {\em et al.}, Submitted to
  Phys. Rev. Lett., hep-ex/0203030.
\bibitem{FOCUSDlife} J. M. Link {\em et al.}, Phys. Lett. B {\bf 537},
  192 (2002). 
\bibitem{SELEXLambdaDzero} A. Kushnirenko {\em et al.},
  Phys. Rev. Lett. {\bf 86}, 5243 (2001).  
\bibitem{CLEODlife} G. Bonvicini {\em et al}, Phys. Rev. Lett. {\bf 82},
  4586 (1999).
\bibitem{PDG} D. E. Groom {\em et. al}, Eur. Phys. J. {\bf 15}, 1 (2000). 
\bibitem{SELEXDs} M. Iori {\em et al.}, Phys. Lett. B {\bf 523}, 22 (2001). 
\bibitem{FOCUSbaryon} J. M. Link {\em et al.}, Phys. Lett. B {\bf
  523}, 53 (2001); J. M. Link {\em et al.}, Phys. Lett. B {\bf
  541}, 211 (2002); J. M. Link {\em et al.}, Phys. Rev. Lett. {\bf
  88}, 161801 (2002). 
\bibitem{CLEObaryon} A. H. Mahmood {\em et al}, Phys. Rev. Lett. {\bf
  86}, 2232 (2001); A. H. Mahmood {\em et al}, Phys. Rev. D {\bf 65},
  031102, (2002). 
\bibitem{D0psi} S. Abachi {\em et al.}, Phys. Rev. Lett {\bf 74}, 2632 (1995).
\bibitem{CDFpsi} F. Abe {\em et al.}, Phys. Rev. Lett {\bf 74}, 2626 (1995).
\bibitem{BraatenCOct}  E. Braaten, S. Fleming, T.C. Yuan, 
Ann. Rev. Nucl. Part. Sci. {\bf 46}, 197 (1996).
\bibitem{babarpromcharmonium} B. Aubert {\em et al.}, Phys. Rev. Lett. {\bf 87}, 162002 (2001).  
\bibitem{bellepromcharmonium} K. Abe {\em et al.}, Phys. Rev. Lett. 
  {\bf 88}, 052001 (2002).
\bibitem{CDFcharm} S. Donati, ``First Run II Results from CDF'',
  XXXVIIth Rencontres de Moriond, Les Arces, 
  March 2002; D. Kaplan, ``Hadron Collider Charm Physics Reach'',
  presented at the Meeting of the Division of Particles and Fields,
  Williamsburg, May 2002. 
\end{thebibliography}
\end{document}